\documentclass{ws-mpla}

\usepackage{latexsym}
\usepackage{graphics}
\newcommand{\bea}{\begin{eqnarray}}
\newcommand{\eea}{\end{eqnarray}}

\begin{document}

\markboth{A. F. Santos}
{G\"{o}del solution in $f(R,T)$ gravity}

\catchline{}{}{}{}{}

\title{G\"{o}del solution in $f(R,T)$ gravity}

\author{\footnotesize A. F. Santos\footnote{alesandroferreira@fisica.ufmt.br}}

\address{Instituto de F\'{\i}sica, Universidade Federal de Mato Grosso,\\
78060-900, Cuiab\'{a}, Mato Grosso, Brazil}

\maketitle

\pub{Received (Day Month Year)}{Revised (Day Month Year)}

\begin{abstract}
In this paper we study G\"{o}del universe in the framework of $f(R,T)$ modified theories of gravity, where $R$ is the curvature scalar and $T$ the trace of the energy momentum tensor. We demonstrate that  G\"{o}del solution occurs in this modified theory and still we suggest a path to understanding the smallness of the cosmological constant.

\end{abstract}

\ccode{PACS Nos.: include PACS Nos.}

\section{Introduction}
The interest to the modified gravity theories is caused by two reasons. The first of them is a quantum one, that is, the problem of constructing the perturbatively consistent gravity theory: it is well known that the usual Einstein gravity is non-renormalizable, whereas, for example, the higher-derivative gravity models such as $R^2$ gravity, suffer from the problem of ghosts, therefore, one needs to develop other models. The second reason is a cosmological one, that is, one must find a consistent gravity theory allowing for explanation of the cosmic acceleration and explain the smallness of the cosmological constant. A number of the theories has been discussed in this context (see for a review \cite{Lima}). One of the new ideas in this context is the generalization of the well-known concept of the $f(R)$ gravity (for a review, see f.e. \cite{rev}, \cite{Clif}, \cite{Capo}) proposing the introduction of a function whose argument is not only the scalar curvature $R$ but also another important scalar, the trace of the energy-momentum tensor of the matter. This concept naturally emerges as a continuation of the idea of $\Lambda(T)$ gravity \cite{Poplawski} which claims that the cosmological constant is a function of the trace of the energy-momentum tensor, $T$. The idea that the Lagrangian is a function of $R$ and $T$ has been proposed in \cite{Harko}, where the equations of motion were derived for the several forms of the function $f(R,T)$, the consistency of the Friedmann-Robertson-Walker (FRW) metric for these forms of $f(R,T)$ was verified within this theory, and the general relativity limit has been discussed. Some cosmological aspects of the $f(R,T)$ gravity have been also discussed in \cite{Chat}, \cite{Mom}, \cite{Chatto} where, in particular, it was shown that the matter compatible with the FRW metric displays a quintom-like behaviour for a certain form of $f(R,T)$. In this context others studies were performed, for example, the laws of the thermodynamics  were verified in \cite{Jardim},  in \cite{Azizi} wormhole solutions were studied, finite-time future singularity was discussed in \cite{Batista} and in \cite{Alvarenga} was shown that the energy conditions, as known in general relativity, can also be applied in this modified theory. The $f(R,T)$ modified theory of gravity is an interesting application of the latter $f(R,L_m)$ gravity proposed in \cite{Lobo}, where the field equations are equivalent to the field equations of the $f(R)$ gravity for empty space-time, but differ from them, as well as from general relativity, when the matter is present.

As the $f(R,T)$ gravity has been studied in several cosmological contexts, the problem of consistency of well-known metrics within this theory naturally arises. One of the first problems in this context is whether the G\"{o}del metric \cite{Godel}, known as a simplest metric allowing for the closed timelike curves (CTCs) is compatible with this theory, or, as is the same, whether the CTCs can be consistent within this theory? This is the problem we address in this paper.

The present paper is organized as follow. In the section 2 we will quickly review the field equations of $f(R, T)$ gravity. The G\"{o}del universe within this modified theory is verified in the section 3. We discuss and conclude in section 4.

\section{Field equations in the $f(R, T)$ gravity}

The action of the $f(R,T)$ gravity model \cite{Harko} is
\bea
\label{ac}
S=\frac{1}{2}\int \sqrt{-g}\left[f(R,T)+2\, L_m\right]d^4x,
\eea
where $f(R,T)$ is an arbitrary function of the scalar curvature $R$, and of the trace $T$ of the energy-momentum tensor of the matter. The dependence from $T$ is investigated as a possible source to introduce exotic fluids or quantum effects (conformal anomaly). The $L_m$ is the matter Lagrangian. In this work, for the sake of the simplicity, we will use the system of units where $8\pi G=1$.  The energy momentum tensor is defined as
\bea
T_{\mu\nu}=-\frac{2}{\sqrt{-g}}\frac{\delta(\sqrt{-g}L_m)}{\delta g^{\mu\nu}},\label{2}
\eea
and its trace is $T=g^{\mu\nu}T_{\mu\nu}$. Assuming that $L_m$ depends only on the components of $g_{\mu\nu}$, and not on their derivatives, and using the well-known properties of the variation with respect to the metric components \cite{inverno}, we get
\bea
T_{\mu\nu}=g_{\mu\nu}L_m-2\frac{\partial L_m}{\partial g^{\mu\nu}}. \label{6}
\eea

First of all, let us briefly discuss the derivation of the equations of motion in this theory.
Varying the action (\ref{ac}) with respect to the metric, we arrive at
\bea
\delta S&=&\frac{1}{2}\int\Bigl[f_R(R,T)\delta R+f_T(R,T)\frac{\delta T}{\delta g^{\mu\nu}}\delta g^{\mu\nu}\nonumber\\
&-&\frac{1}{2}g_{\mu\nu}f(R,T)\delta g^{\mu\nu}+\frac{2}{\sqrt{-g}}\frac{\delta (\sqrt{-g}L_m)}{\delta g^{\mu\nu}}\Bigl]\sqrt{-g}d^4x,\label{Vary}
\eea
where $f_R\equiv \frac{\partial f}{\partial R}$ and $f_T\equiv \frac{\partial f}{\partial T}$.
Since the variation of the scalar curvature is
\bea
\delta R=R_{\mu\nu}\delta g^{\mu\nu}+g_{\mu\nu}\Box\delta g^{\mu\nu}-\nabla_\mu\nabla_\nu\delta g^{\mu\nu},
\eea
and the variation of $T$ can be found as
\bea
\frac{\delta T}{\delta g^{\mu\nu}}=\frac{\delta g^{\alpha\beta}}{\delta g^{\mu\nu}}T_{\alpha\beta}+g^{\alpha\beta}\frac{\delta T_{\alpha\beta}}{\delta g^{\mu\nu}}
=T_{\mu\nu}+ \Theta_{\mu\nu},
\eea
with
\bea
\label{theta}
\Theta_{\mu\nu}\equiv g^{\alpha\beta}\frac{\delta T_{\alpha\beta}}{\delta g^{\mu\nu}},
\eea
we found that the field equations of the $f(R,T)$ gravity model are
\bea
&&f_R(R,T)R_{\mu\nu}-\frac{1}{2}f(R,T)g_{\mu\nu}+(g_{\mu\nu}\Box-\nabla_\mu\nabla_\nu)f_R(R,T)=\nonumber\\
&=&T_{\mu\nu}-
f_T(R,T)(T_{\mu\nu}+\Theta_{\mu\nu}).
\label{14}
\eea

Now let us obtain some relations for the tensor $\Theta_{\mu\nu}$ (\ref{theta}).
Using equation (\ref{6}) one finds
\bea
\frac{\delta T_{\alpha\beta}}{\delta g^{\mu\nu}}=\frac{\delta g_{\alpha\beta}}{\delta g^{\mu\nu}}L_m+\frac{1}{2}g_{\alpha\beta}g_{\mu\nu} L_m-\frac{1}{2}g_{\alpha\beta}T_{\mu\nu}-2\frac{\partial^2L_m}{\partial g^{\alpha\beta}\partial g^{\mu\nu}}.
\eea
Then,
\bea
\Theta_{\mu\nu}=-2T_{\mu\nu}+g_{\mu\nu}L_m-2g^{\alpha\beta}\frac{\partial^2L_m}{\partial g^{\alpha\beta}\partial g^{\mu\nu}}.
\eea

Assuming that the matter Lagrangian is given by $L_m=-p$, as in \cite{Harko}, we rest with
\bea
\Theta_{\mu\nu}=-2T_{\mu\nu}-pg_{\mu\nu}.
\eea

Now let us study the field equations (\ref{14}) for the G\"{o}del universe.

\section{Testing the G\"{o}del universe in the $f(R,T)$ gravity}

Now, let us verify whether the G\"{o}del metric which is written as \cite{Godel}:
\bea
ds^2=a^2\Bigl[dt^2-dx^2+\frac{1}{2}e^{2x}dy^2-dz^2+2 e^x dt\,dy\Bigl],\label{godel}
\eea
where $a$ is a positive number, is compatible with this gravitational model. 
The corresponding non-zero components of the Ricci tensor look like
\bea
R_{00}=1, \,\,\,\,\,\,\,\, R_{02}=R_{20}=e^x, \,\,\,\,\,\,\,\, R_{22}=e^{2x},
\eea
and the scalar curvature is
\bea
R=\frac{1}{a^2}.
\eea

We suppose that the energy-momentum tensor has the same structure as in \cite{Godel}, that is
\bea
T_{\mu\nu}=8\pi\rho u_\mu u_\nu+\Lambda g_{\mu\nu},
\eea
where $u_\mu=(a,0,ae^x,0)$ and $\Lambda$ is the cosmological constant. The non-zero components of the $T_{\mu\nu}$ are
\bea
T_{00}&=&(8\pi\rho+\Lambda)a^2,\nonumber\\
T_{02}&=&(8\pi\rho+\Lambda)a^2e^x,\nonumber\\
T_{11}&=&-\Lambda a^2,\nonumber\\
T_{22}&=&\left(8\pi\rho+\frac{\Lambda}{2}\right)a^2e^{2x},\nonumber\\
T_{33}&=&-\Lambda a^2.
\eea
The trace T of the energy momentum tensor is
\bea
T=8\pi\rho+4\Lambda.\label{trace}
\eea
We will study this compatibility between G\"{o}del metric and $f(R,T)$ gravity for some special cases: 

\subsection{ $f(R,T)=f_1(R)+f_2(T)$}

In this case the field equation becomes
\bea
&&f'_1(R)R_{\mu\nu}-[f_1(R)+f_2(T)]\frac{1}{2}g_{\mu\nu}+(g_{\mu\nu}\Box-\nabla_\mu\nabla_\nu)f'_1(R)\nonumber\\
&=&T_{\mu\nu}-f'_2(T)T_{\mu\nu}-f'_2(T)\Theta_{\mu\nu},
\eea
where the prime denotes a derivative with respect to the argument.

Assuming that matter is a perfect fluid and, for simplicity, we choose the case where the pressure is zero, $p=0$. We have $\Theta_{\mu\nu}=-2T_{\mu\nu}$ and we stay with
\bea
&&f'_1(R)R_{\mu\nu}-\frac{1}{2}g_{\mu\nu}f_1(R)+(g_{\mu\nu}\Box-\nabla_\mu\nabla_\nu)f'_1(R)\nonumber\\
&=&T_{\mu\nu}+f'_2(T)T_{\mu\nu}+\frac{1}{2}g_{\mu\nu}f_2(T).
\eea

We can reformulate this equation as an effective Einstein field equation 
\bea
R_{\mu\nu}-\frac{1}{2}Rg_{\mu\nu}=G_{\rm eff}T_{\mu\nu}+T_{\mu\nu}^{\rm eff},\label{effective}
\eea
where
\bea
G_{\rm eff}&=&\frac{1}{f'_1(R)}\left[1+f'_2(T)\right],\nonumber\\
T_{\mu\nu}^{\rm eff}&=&\frac{1}{f'_1(R)}\Biggl[\frac{1}{2}\Bigl(f_1(R)-Rf'_1(R)+f_2(T)\Bigl)g_{\mu\nu}
-(g_{\mu\nu}\Box-\nabla_\mu\nabla_\nu)f'_1(R)\Biggl].
\eea

Therefore, the gravitational coupling is dependent of the matter.

Making the following choices $f_1(R)=R$ and $f_2(T)=2\lambda T$, where $\lambda$ is a constant we obtain
\bea
G_{\rm eff}&=&1+2\lambda,\nonumber\\
T_{\mu\nu}^{\rm eff}&=&\lambda T g_{\mu\nu}.
\eea
Then the equation (\ref{effective}) becomes
\bea
R_{\mu\nu}-\frac{1}{2}Rg_{\mu\nu}=(1+2\lambda)T_{\mu\nu}+\lambda T g_{\mu\nu}.
\eea

The non-zero components of the motion equations, i.e., (00), (11) and (22) are
\bea
\frac{1}{2}&=&8\pi\rho(1+3\lambda)a^2+\Lambda(1+6\lambda)a^2,\nonumber\\
\frac{1}{2}&=&-8\pi\rho\lambda a^2-\Lambda(1+6\lambda)a^2,\nonumber\\
\frac{3}{4}&=&8\pi\rho\left(1+\frac{5\lambda}{2}\right)a^2+\frac{\Lambda}{2}(1+6\lambda)a^2.
\eea
The components (00) and (02) are the same, as well as (11) and (33). We find as solution these equations

\bea
8\pi\rho&=&\frac{1}{a^2(1+2\lambda)},\nonumber\\
\Lambda&=&-\frac{1+4\lambda}{2a^2(1+2\lambda)(1+6\lambda)}.\label{25}
\eea
If $\lambda = 0$ we recovered the result obtained by G\"{o}del in \cite{Godel}, $8\pi\rho=\frac{1}{a^2}$ and $\Lambda=-\frac{1}{2a^2}$.


\subsection{$f(R,T)=f_1(R)+f_2(R)f_3(T)$}

In this case the field equations (\ref{14}) for the case of a perfect fluid, with $p=0$ (dust), becomes
\bea
&&\left[f'_1(R)+f'_2(R)f_3(T)\right]R_{\mu\nu}-\frac{1}{2}f_1(R)g_{\mu\nu}+(g_{\mu\nu}\Box-\nabla_\mu\nabla_\nu)\left[f'_1(R)+f'_2(R)f_3(T)\right]\nonumber\\
&=&T_{\mu\nu}+f_2(R)f'_3(T)T_{\mu\nu}+\frac{1}{2}f_2(R)f_3(T)g_{\mu\nu}.
\eea
Again, following the suggestions made in \cite{Harko}, i.e., $f_1(R)=R$, $f_2(R)=R$ e $f_3(T)=\lambda T$, we obtain $f'_1(R)=f'_2(R)=1$ e $f'_3(T)=\lambda$. Applying to the G\"{o}del metric, and we assume that the trace of the energy-momentum tensor (\ref{trace}) is a constant, the field equations are reduced to 
\bea
(1+\lambda)R_{\mu\nu}-\frac{1}{2}Rg_{\mu\nu}=T_{\mu\nu}+ R \lambda T_{\mu\nu}+\frac{1}{2}R\lambda T g_{\mu\nu}.
\eea
The non-zero components, i.e., (00), (11) and (22), form the system
\bea
\frac{1}{2}&=&8\pi\rho\left(a^2+\frac{3}{2}\lambda\right)+\Lambda(a^2+3\lambda)-\lambda,\nonumber\\
\frac{1}{2}&=&-8\pi\rho\left(\frac{\lambda}{2}\right) -\Lambda(a^2+3\lambda),\nonumber\\
\frac{3}{4}&=&8\pi\rho\left(a^2+\frac{5\lambda}{4}\right)+\frac{\Lambda}{2}(a^2+3\lambda)-\lambda,
\eea
whereas the component (02) is identical to (00), and (33) to (11). Solving this system, we obtain
\bea
8\pi\rho&=&\frac{1+\lambda}{a^2+\lambda},\nonumber\\
\Lambda&=&-\frac{a^2+\lambda(2+\lambda)}{2(a^2+\lambda)(a^2+3\lambda)}.\label{29}
\eea
If $f_3(T)= 0$ we recovered the result obtained in \cite{Godel}.

\section{Conclusion}

To conclude, we found that in certain cases, i.e., equations (\ref{25}) and (\ref{29}), the G\"{o}del metric solves the equations of motion in the $f(R,T)$ gravity. So, the possibility for arising  the CTCs taking place in a general relativity can be naturally promoted to the $f(R,T)$ gravity. We also observe that the cosmological constant depend on the geometry and matter. In principle, if the parameter $\lambda$ associated with the function $f (T)$ is small, i.e., $\lambda<<1$ and $\lambda<<a$, we recover the usual results of general relativity for the metric G\"{o}del. In the case where $\lambda>>1$ and $\lambda>>a$, we obtain for first case $\Lambda\approx-\frac{1}{6a^2\lambda}$ and for the second case we find a value independent of $\lambda$ for the cosmological constant. Therefore the particular case, $f(R,T)=f_1(R)+f_2(T)$, for a convenient choice of $\lambda$ shows itself interesting because it may indicate a path to understanding the smallness of the cosmological constant.

\section*{Acknowledgements}

This work was supported by the Project FAPEMAT/CNPq No. 685524/2010. The author would like to thank J. R. Nascimento and A. Yu. Petrov for their critical reading of the manuscript, and for helpful discussions.


\begin{thebibliography}{0}
\bibitem{Lima} J. A. S. Lima, {\it Braz. J. Phys.} {\bf 34}, 194 (2004). 

\bibitem{rev} A. De Felice, S. Tsujikawa, {\it Living Rev. Rel.} {\bf 13}, 3 (2010).

\bibitem{Clif} T.~Clifton, P.~G.~Ferreira, A.~Padilla and C.~Skordis, {\it Phys. Rept.} {\bf 513}, 1 (2012).

\bibitem{Capo} S. Capozziello, M. De Laurentis, {\it Phys. Rept.} {\bf 509}, 167 (2011).   
 
\bibitem{Harko} T. Harko, F. S. N. Lobo, S. Nojiri and S. D. Odintsov, {\it Phys. Rev. D} {\bf 84}, 024020 (2011).

\bibitem{Poplawski} N. J. Poplawski, arXiv:  gr-qc/0608031.

\bibitem{Chat} M. S. J. Houndjo, {\it Int. J. Mod. Phys. D} {\bf 21},1250003 (2012).

\bibitem{Mom} D. Momeni, R. Jamil, R. Myrzakulov, {\it Eur. Phys. J. C} {\bf 72}, 1999 (2012).

\bibitem{Chatto} S. Chattopadhyay, arXiv: 1208.3896.

\bibitem{Jardim} M. J. S. Houndjoa, F. G. Alvarenga, M. E. Rodrigues  and D. F. Jardim, arXiv: 1207.1646.

\bibitem{Azizi} T. Azizi, arXiv: 1205.6957.

\bibitem{Batista} M. J. S. Houndjo, et al, arXiv: 1203.6084.

\bibitem{Alvarenga}F. G. Alvarenga, et al, {\it Journal of Modern Physics} {\bf 4}, 130 (2013).

\bibitem{Lobo} T. Harko and F. S. N. Lobo, {\it Eur. Phys. J. C} {\bf 70}, 373 (2010).

\bibitem{Godel} K. G\"{o}del, {\it Rev. Mod. Phys.} {\bf 21}, 447 (1949).

\bibitem{inverno} R. D'Inverno, {\it Introducing Einstein's Relativity}, (Oxford University Press, 1992) 145-152.
\end{thebibliography}
\end{document}